\documentclass[aps,a4paper,10pt,twocolumn,nofootinbib]{revtex4} 
\usepackage{amsfonts}
\usepackage{amssymb}
\usepackage{mathrsfs}
\usepackage[mathscr]{euscript}
\usepackage[dvips]{graphicx}
\usepackage[hypertex=true]{hyperref}
\usepackage{fancyhdr}
\usepackage{colordvi}

\begin{document}

\title{Optical carrier wave shocking: detection and dispersion}
\author{P. Kinsler}
\affiliation{
  Blackett Laboratory, Imperial College London,
  Prince Consort Road,
  London SW7 2BW, 
  United Kingdom.}
\email{Dr.Paul.Kinsler@physics.org}
\author{S.B.P. Radnor}
\affiliation{
  Blackett Laboratory, Imperial College London,
  Prince Consort Road,
  London SW7 2BW, 
  United Kingdom.}
\author{J.C.A. Tyrrell}
\affiliation{
  Blackett Laboratory, Imperial College London,
  Prince Consort Road,
  London SW7 2BW, 
  United Kingdom.}
\author{G.H.C. New}
\affiliation{
  Blackett Laboratory, Imperial College London,
  Prince Consort Road,
  London SW7 2BW, 
  United Kingdom.}

\begin{abstract}

Carrier wave shocking is studied using 
 the Pseudo-Spectral Spatial Domain (PSSD) technique.  
We describe the shock detection diagnostics necessary 
 for this numerical study, 
 and verify them against theoretical shocking predictions 
 for the dispersionless case.
These predictions show 
 Carrier Envelope Phase (CEP) and 
 pulse bandwidth sensitivity  
 in the single-cycle regime. 
The flexible dispersion management offered by PSSD
 enables us to independently control 
 the linear and nonlinear dispersion.
Customized dispersion profiles
 allow us to analyze the development of 
 both carrier self-steepening and shocks.
The results exhibit a marked 
 asymmetry between normal and anomalous dispersion,
 both in the limits of the shocking regime
 and in the (near) shocked pulse waveforms. 
Combining these insights, 
 we offer some suggestions on how
 carrier shocking (or at least extreme self-steepening) 
 might be realised experimentally.

\end{abstract}

\pacs{X}


\newcommand{\sech}{{\textrm{ sech}}}

\lhead{\includegraphics[height=5mm,angle=0]{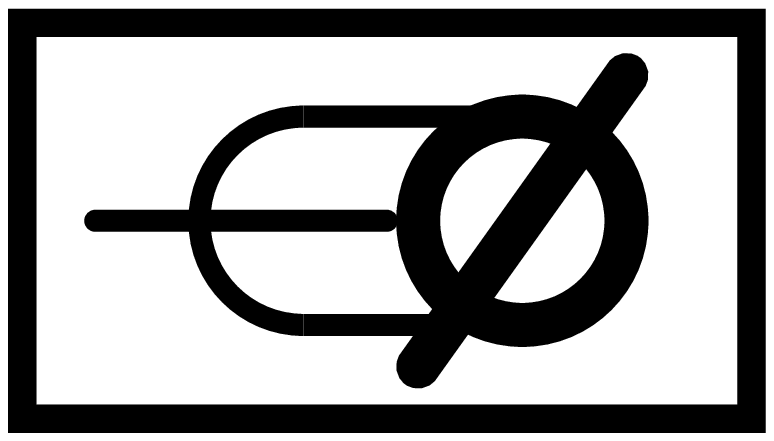}~~CSHOCK-II}
\chead{~}
\rhead{
\href{mailto:Dr.Paul.Kinsler@physics.org}{Dr.Paul.Kinsler@physics.org}
}
\lfoot{} 
\rfoot{Kinsler-RTN-2007} 

\date{\today}
\maketitle
\thispagestyle{fancy}

%

%
\section{Introduction}\label{S-introduction}

The self-steepening of an optical pulse envelope
 was first studied by DeMartini et al. 
 in 1967 \cite{DeMartini-TGK-1967pr},
 and is a well-known phenomenon associated with self-phase modulation (SPM).
Surprisingly however, 
 the possibility of self-steepening of the 
   optical carrier wave\footnote{
Note that in this paper we use the term ``carrier'' to denote 
 all the oscillations of the field, 
 and do not use its other sense, 
 i.e.  that of fixed-frequency oscillations 
 as used in a envelope and carrier representation of a pulse.
}
 was considered even earlier
 in a 1965 paper by Rosen \cite{Rosen-1965pr},
 who showed that, 
 for a third order $\chi^{(3)}$ nonlinearity 
 (and under suitable conditions), 
 a shock (or field discontinuity)
 develops in a finite distance.
This latter phenomenon received little
 attention for more than 30 years, 
 until it was revisited in the 1990s by
 Moloney et al. \cite{Flesch-PM-1996prl,Gilles-MV-1999pre},
 who performed Finite Difference Time Domain (FDTD) simulations 
 of the process.
Three dimensional FDTD simulations of carrier shocking have recently
 been performed by Trillo et al. \cite{Trillo-MC-2006ecleo}.

In the present paper,
 we investigate carrier wave shocking 
 in $\chi^{(3)}$ nonlinear materials where 
 SPM is accompanied by 
 the generation of (odd) higher harmonics.
For the dispersionless case we have
 generalized earlier predictions based on 
 the method of characteristics (MOC)
 to allow for arbitrary initial waveforms.
This allows us, 
 for example, 
 to predict the carrier envelope phase (CEP) 
 sensitivity of the shocking distance for optical pulses, 
 as well as the dependence on pulse length.

Our primary interest in this paper is the effect of linear dispersion  
 on carrier shock formation.
Since no analytic solutions exist in this case, 
 we are forced to rely on numerical simulations.
In any discussion of carrier wave shocking, 
 it is important to distinguish between three related concepts: 
 the physical system, 
 the mathematical model, 
 and the numerical model.  
The mathematical model is an approximation to the physical system, 
 while the numerical model is an approximation to the mathematical model.   
Actual discontinuities occur only in the mathematical model, 
 and it is these to which the idea of shock formation refers.  
In the physical system, 
 discontinuities are prevented by phenomena not included in the mathematics, 
 while numerical codes inevitably fail 
 as a mathematical discontinuity is approached.  
The indicator of an imminent ``shock'' is 
 the rapidly increasing gradient (steepening) of the optical carrier. 
In the numerical model, 
 we see numerical symptoms generated by 
 extreme self-steepening, 
 and these correlate with the onset of the discontinuities
 in the mathematical model.
Under these circumstances, 
 it has been necessary for us to develop a quantitative numerical test 
 of ``shock formation''.  
We have found the most satisfactory diagnostic to be
 ``Local Discontinuity Detection'' (LDD), 
 which gives results that are in good agreement with 
 theoretical MOC predictions based on the mathematical model.  
LDD provides a clear numerical measure of 
 the rapid steepening that precedes the appearance
 of a discontinuity in the mathematical representation. 
We continue to use the term ``shock'' to describe 
 the situation where a mathematical
 shock is imminent.   

In our simulations, 
 we exploit the flexibility in dispersion management offered by the 
 Pseudo-Spectral Spatial Domain (PSSD) technique \cite{Tyrrell-KN-2005jmo}
 to study carrier shock formation for a range of 
 simple dispersion profiles, 
 and determine the degree of phase mismatch that 
 carrier shocking can tolerate.
As expected, 
 we find that shocking occurs when the nonlinearity 
 dominates the linear dispersion. 
However,
 it emerges that the process is asymmetric,
 with anomalous dispersion being 
 far more conducive to shocking than normal dispersion.  
Hence, 
 the fact that the LDD scheme does not detect carrier shocks 
 in simulations involving (normally dispersive) fused silica,
 even at powers equal to its damage threshold, 
 is neither surprising nor necessarily discouraging.  
After all, 
 anomalously dispersive materials could potentially be engineered.  
Further, 
 our results also relate to how one might perform carrier shaping 
 (as opposed to carrier steepening), 
 a process that has some interesting applications.

After briefly describing our simulation methods in 
 section \ref{S-numericals}, 
 we consider dispersionless shocking and 
 the method of characteristics in \ref{S-dispersionless}, 
 and
 our LDD shock detection scheme in \ref{S-shockdetection}.
Then we discuss 
 the effect of dispersion on carrier shocking in \ref{S-dispersive}, 
 followed by our numerical results in \ref{S-results}.
In 
 \ref{S-applications} we consider the potential relevance
 to experimental detection of carrier steepening and/or shocks.
Finally,
 in \ref{S-conclusions},
  we present our conclusions.

%
\section{Simulation Methods}
\label{S-numericals}

The PSSD method \cite{Tyrrell-KN-2005jmo,Fornberg-PSmethods}
 offers significant advantages over the traditional FDTD
 and Pseudospectral Time-Domain (PSTD) \cite{Hagness} techniques
 for modeling the propagation and interaction of few-cycle pulses.
Run times are generally faster, 
 and PSSD also offers far greater flexibility 
 in the handling of dispersion.
Whereas FDTD and PSTD \cite{Hagness} propagate fields $E(z), H(z)$ 
 forward in {\em time}, 
 PSSD propagates fields $E(t), H(t)$ forward in {\em space}.
It is important to keep this difference in mind when comparing 
 our results to those in \cite{Hagness}.
Under PSSD, the entire time-history (and therefore frequency content)
 of the pulse is known at any point in space, 
 so arbitrary dispersion incurs no extra computational penalty.
In contrast, 
 the FDTD or PSTD approaches use convolutions 
 incorporating time-response models for dispersion.

We apply the PSSD algorithm to two representations of the field and 
 source-free Maxwell's equations in non-magnetic media;
 the first uses the $E$ and $H$ fields,
 and the second the directional fields
 $\vec{G}^{\pm} (t)
~=~
  \alpha_r (t)  \ast \vec{E}_x  (t)
 \pm 
  \beta_r \vec{H}_y (t)$ 
 \cite{Kinsler-RN-2005-flvar}.
Here the 
$\alpha_r$, $\beta_r$ include 
 the (linear) permittivity and permeability of the material 
 (i.e. $\epsilon(t), \mu$).
These $G^\pm$ fields
 enable us to rewrite Maxwell's equations, 
 and efficiently separate out the relevant 
 forward-going part of the field.

For an instantaneous $\chi^{(3)}$ nonlinearity, 
 the equations for $E$ and $H$ in the 1D (plane wave) limit are
~
\begin{eqnarray}
  \frac{dH_y(t;z)}{dz} 
&=&
 -
  \frac{d}{dt}
  \left[
    \epsilon_0 \epsilon_r (t) * E_x(t;z) + \chi^{(3)}  E_x(t;z)^3
  \right]
,
\label{eqn-pssd-dH}
\\
  \frac{dE_x(t;z)}{dz} 
&=&
 -
  \frac{d}{dt}
  \left[
    \mu_0 H_y(t;z)
  \right]
.
\label{eqn-pssd-dE}
\end{eqnarray}
The $G^\pm$ field simulations usually assume $G^-=0$, 
 and as a result contain only forward traveling components.
The forward-only wave equation for $G^+$ is
~
\begin{eqnarray}
  \frac{dG^+(t;z)}{dz} 
&=&
 -
  \frac{d}{dt}
  \left[
    \beta_r \alpha_r (t) * G^+(t;z) + \beta_r \chi^{(3)} E(t;z)^3
  \right]
,
\label{eqn-pssd-dG}
\end{eqnarray}
where it is most straightforward to calculate the nonlinear term by 
 reconstructing $E(t;z)$ from $G^+(t;z)$ 
 in the frequency domain using $\tilde{E}(\omega;z) 
  = \tilde{G}^+(\omega;z) / 2 \tilde{\alpha}_r(\omega)$, 
 since $G^-=0$.
Notice the similarity between eqns. (\ref{eqn-pssd-dH}) 
 and (\ref{eqn-pssd-dG}), 
 but that eqn. (\ref{eqn-pssd-dG}) propagates 
 the field in a single first order equation, 
 rather than two.

Typical array sizes used in pulse simulations were $N=2^{14}$
 covering a time window $T=200$fs,
 and (spatial) propagation steps were $dz = 0.4 c T/N \approx 0.9$nm. 
We ensured the stability of our integration using 
 Orszag's 2/3 rule \cite{NumericalRecipes},
 which involves setting the upper part of the spectral range to zero.
It is worth noting that changing this cut-off, 
 either by adjusting its position, 
 or by using a smoothed (rather than step-like) filter,
 made little difference to test simulations.
The pulse profile used as an initial condition was 
~
\begin{eqnarray}
  E(t) 
&=&
  E_0 
  \sin(\omega_1 t + \phi) 
  \sech( 0.28 \omega_1 t / \tau)
,
\label{eqn-initialfield}
\end{eqnarray}
where our standard parameters were $\omega_1=2.356\times 10^{15}$rad/s 
 (i.e. $\lambda = 800$nm) with $\tau = 0.9\dot{3}$.
Such pulses are rather short
 (since the number of cycles inside the intensity FWHM is $\tau$), 
 but in fact the shocking distance is only weakly dependent on 
 the pulse width, 
 with significant variation 
 only appearing for pulses of a few cycles or less.

We also performed CW simulations, 
 for which we modeled just a single cycle of the carrier.
 assisted by the periodic nature of the discrete Fourier transform.
For these we used array sizes of $N=2^{10}$, 
 and the time window was set by the period of the field oscillations.

Our default value of nonlinear strength was $\chi^{(3)} E_0^2 = 0.02$, 
 which is comparable to that in
 fused silica at an intensity of $ 0.7 \times 10^{14}$ W/cm$^2$;
 our $\chi^{(3)}$ parameter is equivalent to 
 $\eta$ in Rosen \cite{Rosen-1965pr}, 
 and
 $a$ in Gilles et al.\cite{Gilles-MV-1999pre}.
We use an instantaneous nonlinearity, 
 since our primary interest is linear dispersion, 
 and that is a far more significant effect than 
 the nonlinear time response.




%
\section{Carrier Wave Shocking}
\label{S-dispersionless}

\begin{figure}[ht]
\includegraphics[height=0.80\columnwidth,angle=-90]{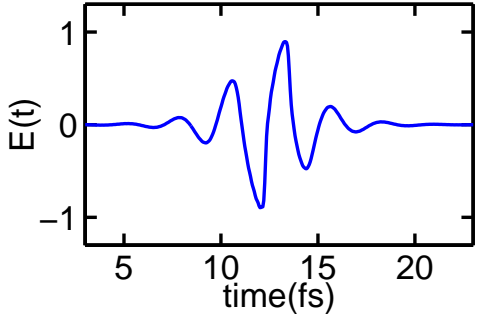}
\caption{
The profile of a few-cycle optical pulse 
 just prior to shocking in the dispersionless limit.
The larger oscillations in the centre of the pulse undergo more 
 self-steepening than those in the wings.
The standard pulse parameters were used.
}
\label{fig-pulsecomparison-t0}
\end{figure}

As an introduction to the process of carrier wave shocking,
 fig. \ref{fig-pulsecomparison-t0} shows 
 the profile of a pulse propagating in a dispersionless 
 $\chi^{(3)}$ medium, 
 just before a shock occurs. 
The nonlinearity gives rise to a
 nonlinear index of refraction $n_2 E^2$, 
 the effect of which is
 to increase the effective refractive index
 in the more intense regions of the profile.
This 
 reduces the phase velocity at the peak of each oscillation 
 with respect to the rest of the waveform, 
 and causes 
 the slope on the trailing edges to increase dramatically.

\begin{figure}[ht]
\begin{center}
\includegraphics[height=0.80\columnwidth,angle=-90]{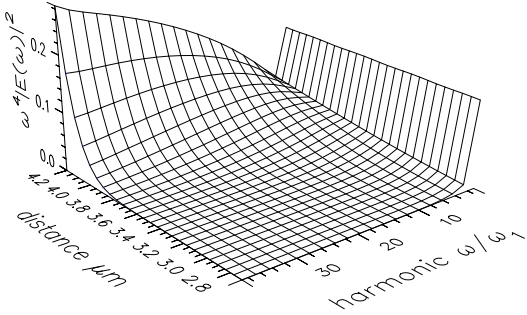}
\end{center}
\caption{
Development of the heights of the scaled harmonic peaks
 as a pulse approaches the (LDD) shocking distance of 4.3$\mu$m
 in the dispersionless case.
Each line perpendicular to the 
 $\omega/\omega_1$
 axis 
 corresponds to the (scaled) contribution from that spectral peak.
Note that the viewpoint has been rotated so that the 
 contribution from the fundamental is to the {\em right}.
The initial pulse contained about 33 cycles ($\tau=33$).
}
\label{fig-pulsecomparison-w}
\end{figure}

The effects seen in fig. \ref{fig-pulsecomparison-t0} are
 associated with the generation of third and higher harmonics,
 although the harmonic components are not
 particularly strong even when a shock is about to occur.  
As an example, 
 fig. \ref{fig-pulsecomparison-w} shows
 how the harmonics build up as shocking is approached.  
The $\omega^4$ scaling of the intensity spectrum 
 in the figure exaggerates the contribution of the higher orders, 
 and has been chosen for illustrative purposes.  
Notice that the profile becomes nearly flat 
 (i.e. the spectrum falls off as the 4th power of the frequency) 
 just before shocking is registered at around 4.3$\mu$m.

If we write 
~
\begin{eqnarray}
  E 
&=&
  A(t) 
  \left[
    \sin(\omega_1 t) + \gamma \cos(3 \omega_1 t + \psi)
  \right]
, 
\label{eqn-Eplus3H}
\end{eqnarray}
 we find that an appropriate choice of $A(t)$, 
 with $\gamma=0.1$ and $\psi=0$,  
 gives us a passable match to fig. \ref{fig-pulsecomparison-t0}.
Note that this choice of $\psi$ corresponds to the phase
 of third harmonic generation (THG) under index matched conditions, 
 i.e. where $n_0 = n(\omega_1)=n(\omega_3)$.

Rosen's original paper \cite{Rosen-1965pr}
 used the MOC to 
 predict the formation of a value discontinuity in the field
 at certain points within the profile.
If the displacement of the dispersionless medium is written
~
\begin{eqnarray}
 D 
&=&
  \epsilon_0 \left( E + \chi^{(1)} E + \chi^{(3)} E^3 \right)
,
\label{eqn-Dexpansion}
\end{eqnarray}
he showed that the wave equation for $E$ is 
\begin{eqnarray}
  c^2 \frac{\partial^2 E} {\partial x^2}
&=&
  \left(1 + \chi^{(1)}\right) 
  \frac{\partial^2 E }
       {\partial t^2}
 + \chi^{(3)} 
   \frac{\partial^2 E^3} 
        {\partial t^2}
,
\label{eqn-Ewaveeqn}
\end{eqnarray}
 and that the associated equation 
 governing the characteristic lines of $E$ is
~
\begin{eqnarray}
  \frac{\partial E} 
       {\partial t} 
 + v ( E ) 
   \frac{\partial E} 
        {\partial x}
&=&
  0.
\label {eqn-characteristcs}
\end{eqnarray}
Here, 
 the velocity $v(E)$ is given by
~
\begin{eqnarray}
  v(E) 
&=&
  \frac {c} {(\epsilon_r + 3 \chi^{(3)} E^2)^{1/2}}
,
\label{eqn-vE}
\end{eqnarray}
where $\epsilon_r = 1 + \chi^{(1)} = {n_0}^2$ is 
 the (relative) dielectric constant 
 and $n_0$ the linear refractive index.

\begin{figure}[ht]
\includegraphics[width=0.60\columnwidth,angle=-0]{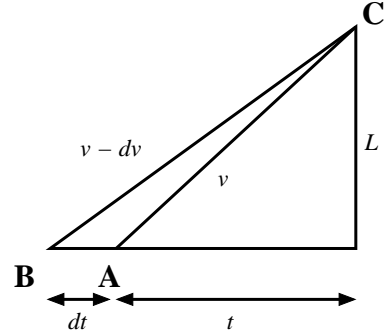}
\caption{
Method of Characteristics.
Two points A and B on the field profile, 
 separated initially by a time difference $dt$ travel
 at different speeds $v-dv$ and $v$, 
 and meet at point C.
}
\label{fig-moc}
\end{figure}

Using eqn. (\ref{eqn-vE}) along with 
 the construction shown in fig. \ref{fig-moc}, 
 we can derive a simple formula for 
 the distance to shocking. 
The figure shows two characteristics AC and BC, 
 originating from points A and B, 
 and converging towards a shock at C after a distance of $L$.  
The intensity at A is higher than at B, 
 so the speed associated with AC (represented by its gradient)
 is lower than that of BC.  
This means that at C, 
 the field has two values, 
 and a discontinuity has formed.
From the geometry of the figure, 
 it is easy to show that
~
\begin{eqnarray}
  \frac{dv}{dt}
&=&
  \frac{v}{t}
=
  \frac{v^{2}}{L}
=
  \frac{(c/n_0)^{2}}{L},
\label{eqn-moc1}
\end{eqnarray}
where $t$, 
 $v=c/n_0$, 
 and $L=vt$ are respectively time, 
 speed and distance.

On the other hand, 
 differentiating eqn. (\ref{eqn-vE}) leads to
~
\begin{eqnarray}
  \frac{dv}{dt}
&=&
  -
  \frac{3 c \chi^{(3)}} 
  {2 \left( n_o^{2} + 3\chi^{(3)} E^2\right)^{3/2} }
  \frac{d\left(E^2\right)}{dt}
,
\label{eqn-moc2}
\end{eqnarray}
and this combined with eqn. (\ref{eqn-moc1})
 yields
~
\begin{eqnarray}
  L 
&=&
  \frac{c n_0 \sqrt{1 + 3\chi^{(3)} E^2/n_o^2}}
       {3 \chi^{(3)} (-dE^2/dt)}
\\
&=&
  \frac{c}
       {4 n_2}
  \frac{\sqrt{1 + 8 n_2 E^2/n_o}}
       {(-dE^2/dt)}
,
\label{eqn-moc3}
\end{eqnarray}
 where $n_2 = 3 \chi^{(3)} / 8 n_o$
 is the material parameter determining the 
 intensity induced refractive index shift as $n_2 E^2$.
For a given profile, 
 a shock will occur first at the point where
 $-dE^2/dt$ reaches its negative extremum.
We can therefore define the shocking distance as
~
\begin{eqnarray}
  L_{shock}
&=&
  \frac{c }
       {4 n_2}
  Min{\frac{\sqrt{1 + 8 n_2 E^2/n_o} }{ (-dE^2/dt)}}
\\
&\simeq&
  \frac{c}
       {4 n_2}
  Min{\frac{1}{(-dE^2/dt)}}
~~~~
\textrm{for}
~~~~
  8 n_2 E^2/n_o \ll 1
.
~~~~
\label{eqn-moc4}
\end{eqnarray}
This formula is more general than that of either Rosen \cite{Rosen-1965pr}
 or Gilles et al. \cite{Gilles-MV-1999pre}.
Notice in particular that the parameter 
 that controls the shock behaviour
 is not the gradient of the field ($dE/dt$), 
 but that of the field squared ($dE^2/dt$).

For a sinusoidal initial waveform 
 $E =E_{0}\sin (\omega_1 t + \phi)$,
 it is easy to show that shocks form at
~
\begin{eqnarray}
  \omega_1 t 
&=& 
  - \pi / 4 + j \pi - \phi
,
\label{eqn-location}
\end{eqnarray} 
where $j$ is an integer, 
 and that the shocking distance is 
~
\begin{eqnarray}
  L_{shock}
&=&
  \frac{2 c n_0}{3 \omega_1 \chi^{(3)} E_{0}^{2}}
~~~~
=
  \frac{c}{4 \omega_1 n_2 E_{0}^{2}}
.
\label{eqn-moc5}
\end{eqnarray}
The approach is readily extended to pulsed waveforms of the type defined
 in eqn. (\ref{eqn-initialfield}).  
Analytical results can be derived in the new situation
 in which
 eqn. (\ref{eqn-location}) becomes a transcendental equation.  
However, 
 the results are cumbersome and it is simpler
 to scan the profile numerically to determine the shocking parameters.  
It turns out that
 the shocking distance for short pulses exhibits an interesting
 sensitivity to the carrier envelope phase $\phi$.  
Whilst in the CW case
 all locations defined by eqn. (\ref{eqn-location}) were equivalent, 
 for pulses, 
 the one nearest the peak of the envelope 
 has a shorter $L_{shock}$ than the others.

\begin{figure}[ht]
\begin{center}
 \includegraphics[height=0.80\columnwidth,angle=-90]{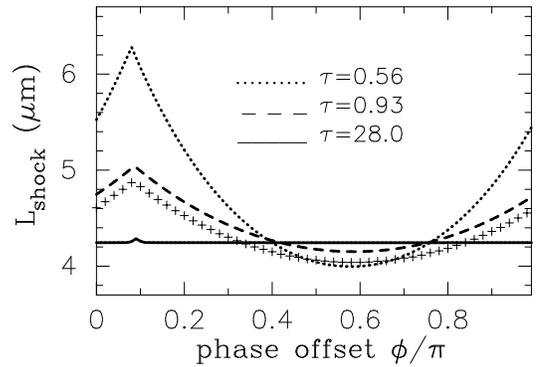}
\end{center}
\caption{
MOC shocking distances as a function of phase, 
 for pulses as in eqn. (\ref{eqn-initialfield}), 
 allowing for different pulse lengths. 
We include the 1/2 cycle $\tau=0.56$ results
 to emphasize the trend; 
 but even for a long pulse ($\tau=28$),
 a small peak can be seen just below $\phi/\pi = 0.01$.
The peak position is weakly $\tau$ dependent.
The $+$ signs denote LDD shocking distances obtained from simulations
 of the $\tau=0.9\dot{3}$ case.
}
\label{fig-ceo}
\end{figure}

A set of results is displayed in fig. \ref{fig-ceo} 
 where the shocking distance
 is plotted as a function of 
 the carrier phase $\phi$ for $\sech$ profiles with 
 different pulse widths $\tau$.  
The dotted line is for the case of a very broad pulse for which 
 $L_{shock}$ is given by eqn. (\ref{eqn-moc5}).  
The sharp peaks mark a curve crossings where the shock location switches from 
 one point on the sinusoid to another.   
Notice that the range of shocking distances increases for shorter pulses, 
 and that, 
 unsurprisingly, 
 the lowest values occur when $\phi$ is around $\pi/2$ 
 i.e. when the pulse has a cosine form.

One very important point to note from fig. \ref{fig-ceo} is 
 that for pulses containing more than a few cycles, 
 the dependence of the shocking distance on pulse width is
 very weak.
We will see this message repeated later 
 in section \ref{S-dispersive}, 
 with shock regions being similar for both single cycle ($\tau=1$) pulses
 and CW ($\tau=\infty$) fields.

\section{Shock Detection}
\label{S-shockdetection}

Since optical shock formation is directly associated with 
 regions of increasingly steep field gradient, 
 any numerical scheme, 
 however sophisticated, 
 is bound to fail at some point in the process.
We therefore want to recognize when 
 a shock is imminent,
 not only to avoid numerical problems, 
 but as a means to estimate the 
 distance at which a discontinuity would occur in the 
 mathematical model.

One obvious symptom of impending numerical failure is loss
 of energy conservation \cite{Tyrrell-KN-2005jmo}.
However,  
 this does not give an indication 
 at the first instance 
 of a shock forming, 
 but rather signals the accumulated effect of multiple 
 small numerical failures from 
 many shocked regions.

A more physical strategy is to search for regions
 where the field gradient $dE/dt$ is large and increasing rapidly, 
 and to use this to predict the shocking point.
A useful variant, 
 suggested by the MOC calculation
 in the previous section, 
 is to use the value of $-dE^2/dt$ instead of the gradient.

\begin{figure}[ht]
\begin{center}
 \includegraphics[width=0.50\columnwidth,height=0.80\columnwidth,angle=-90]{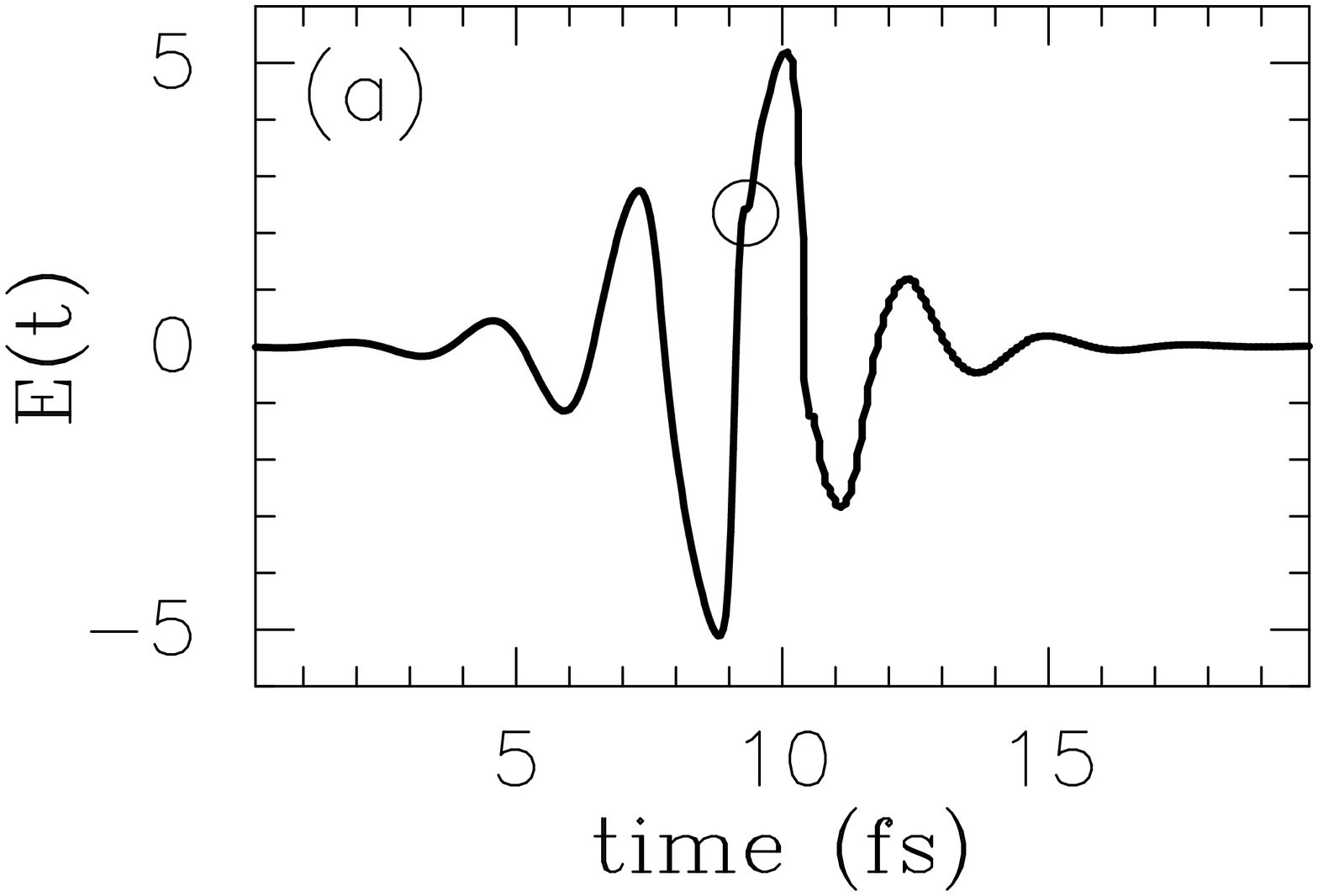}\\
 \includegraphics[width=0.50\columnwidth,height=0.80\columnwidth,angle=-90]{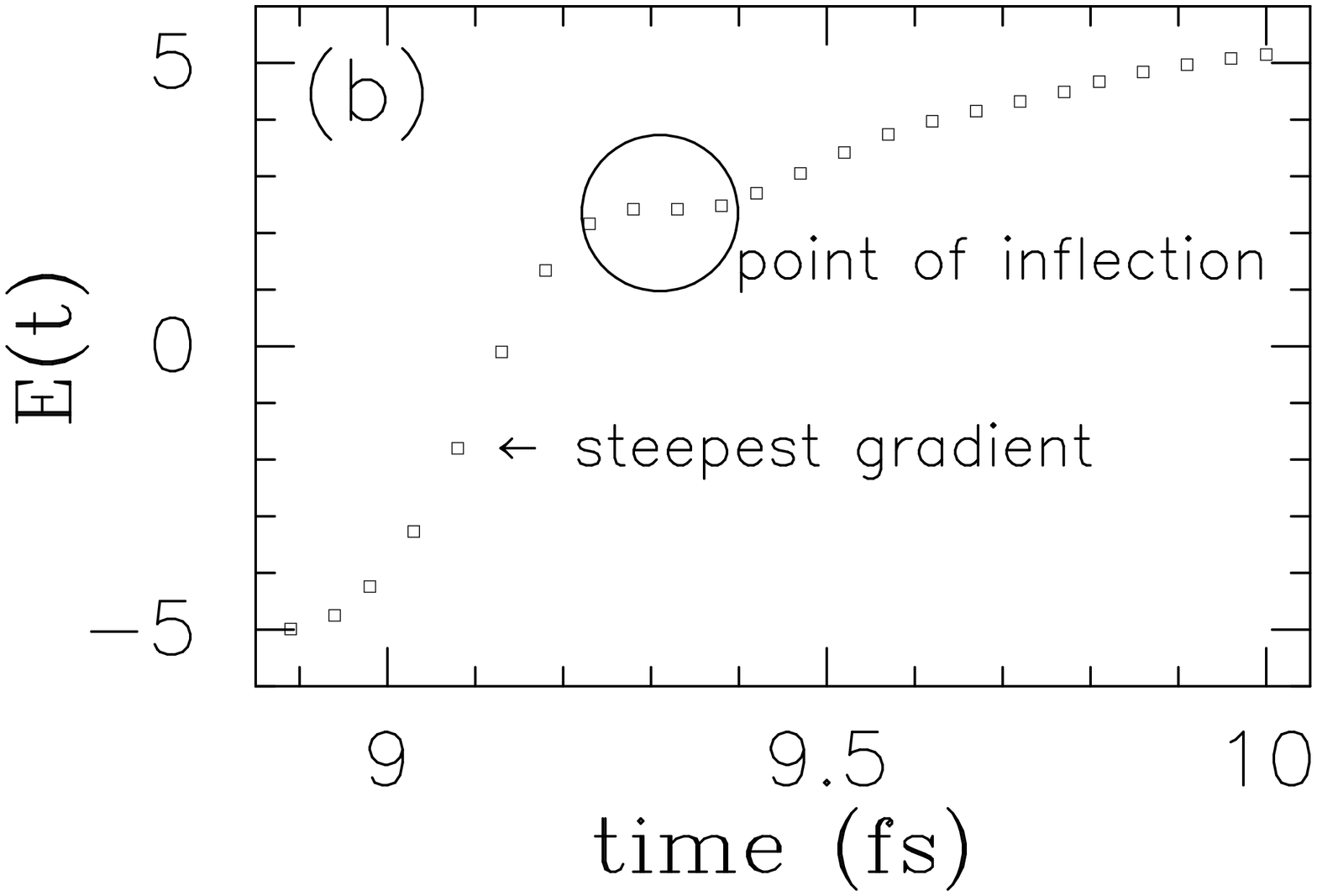}
\end{center}
\caption{
The profile of a few-cycle optical pulse 
 at the point of shocking.
The larger oscillations in the centre of the pulse undergo more 
 self-steepening than those in the wings.
(a) The whole pulse, 
 with the LDD carrier shock circled ($\circ$).
(b) An expanded view of the shock region, 
with its point of inflection (at $\sim 9.3$fs) being very close to  
the steepest gradient (at $\sim 9.1$fs).
The standard pulse parameters were used.
}
\label{fig-expandedradlet}
\end{figure}

Overall, 
 we find that the best method is Local Discontinuity Detection (LDD), 
 which is similar to techniques used in other fields
 (see e.g. \cite{PagendarmSeitz-1993}).
As the shock regime is approached, 
 narrow shoulders with associated 
 points of inflection appear
 within the regions of rapidly increasing gradient.
The procedure is therefore to scan the field profile for 
 the maximum gradient (of either $E$ or $E^2$) and, 
 if it occurs near a point of inflection, 
 an incipient shock is registered.
An example of a pulse that has just triggered the LDD
 diagnostic is shown in
 fig \ref{fig-expandedradlet}.

The LDD method requires two parameters. 
The first determines the time scale used in 
 determining whether a point of inflection exists.
For this, 
 we pick the scale set by our temporal grid and 
 insist that the field gradients calculated at three adjacent 
 grid points have opposing signs:
 either up-down-up, 
 or down-up-down.
The second determines the maximum range allowed between the 
 maximum gradient and the point of inflection, 
 and our default value for this was 10 grid points.
In our simulations, 
 we see that the position of the first detected shock 
 depends only weakly on this range.
We can easily minimize the small sensitivity to 
 these parameters by holding them fixed throughout any given 
 set of simulations.
As a result, 
 we have found LDD to be a sensitive and reliable method of shock detection.

\begin{figure}[ht]
\begin{center}
 \includegraphics[height=0.90\columnwidth,angle=-90]{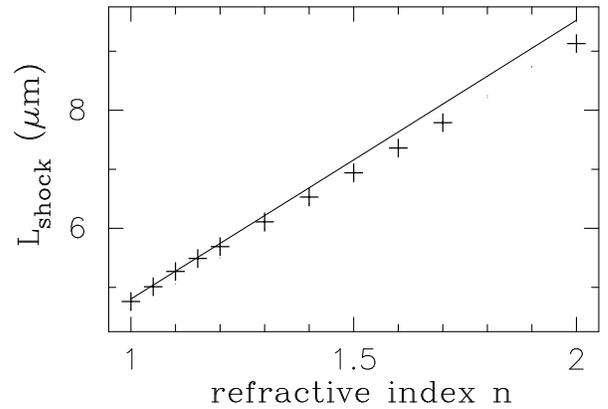}
\end{center}
\caption{
The LDD shocking distance as a function of 
 refractive index, 
 comparing the approximate MOC prediction from 
 eqn. (\ref{eqn-moc4}) 
 to PSSD simulations ($+$) for pulses with $\tau=0.9\dot{3}$.
Other PSSD simulation results from independent codes
 give very similar results to those 
 shown on the graph.
}
\label{Ls-vs-n}
\end{figure}

Eqn. (\ref{eqn-moc5}) predicts that the shocking distance should 
 increase linearly with the refractive index $n_0$. 
We use this to test the LDD diagnostic 
 in fig. \ref{Ls-vs-n},
 where the analytical formula is compared with the 
 results of numerical simulations.
This figure shows close agreement between prediction 
 (using eqn. (\ref{eqn-moc4})) and simulation, 
 where the approximation causes the MOC prediction
 to be reduced by less than 0.1$\mu$m at $n_0=1$.
We see similar agreement between simulations using 
 the LDD method and the carrier phase sensitivity
 shown in fig. \ref{fig-ceo}.
The presence of small systematic differences
 (as on e.g. fig. \ref{fig-ceo}, \ref{Ls-vs-n}) can be easily understood, 
 since the LDD diagnostic is (strictly speaking) a test of the numerics, 
 and is not a direct test for the presence of a physical shock
 or mathematical discontinuity.

%
\section{The effect of dispersion on shocking }
\label{S-dispersive}

%
%
%

\begin{figure}[ht]
\includegraphics[height=0.90\columnwidth,angle=-90]{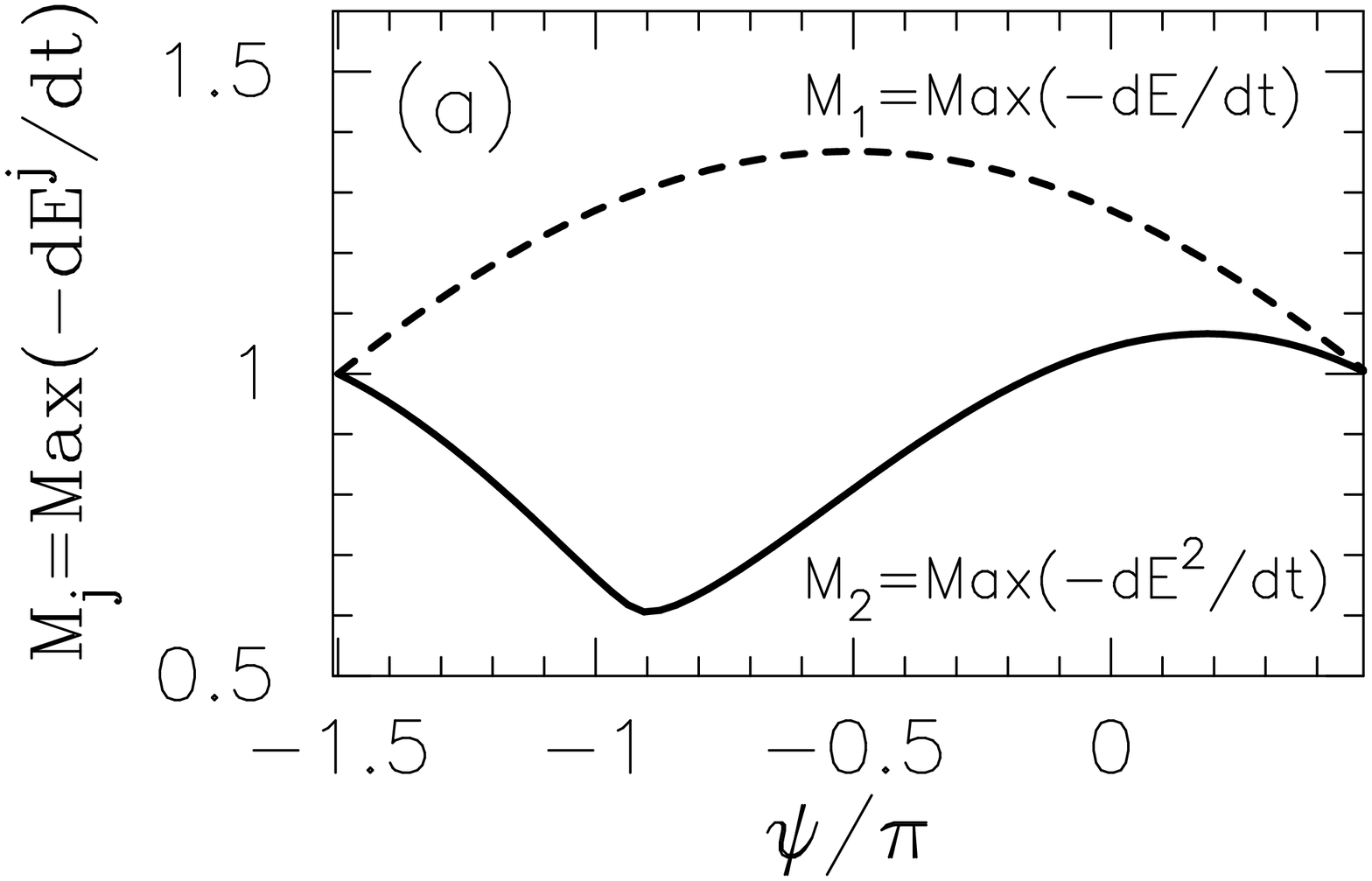}\\
~\\
\includegraphics[height=0.90\columnwidth,angle=-90]{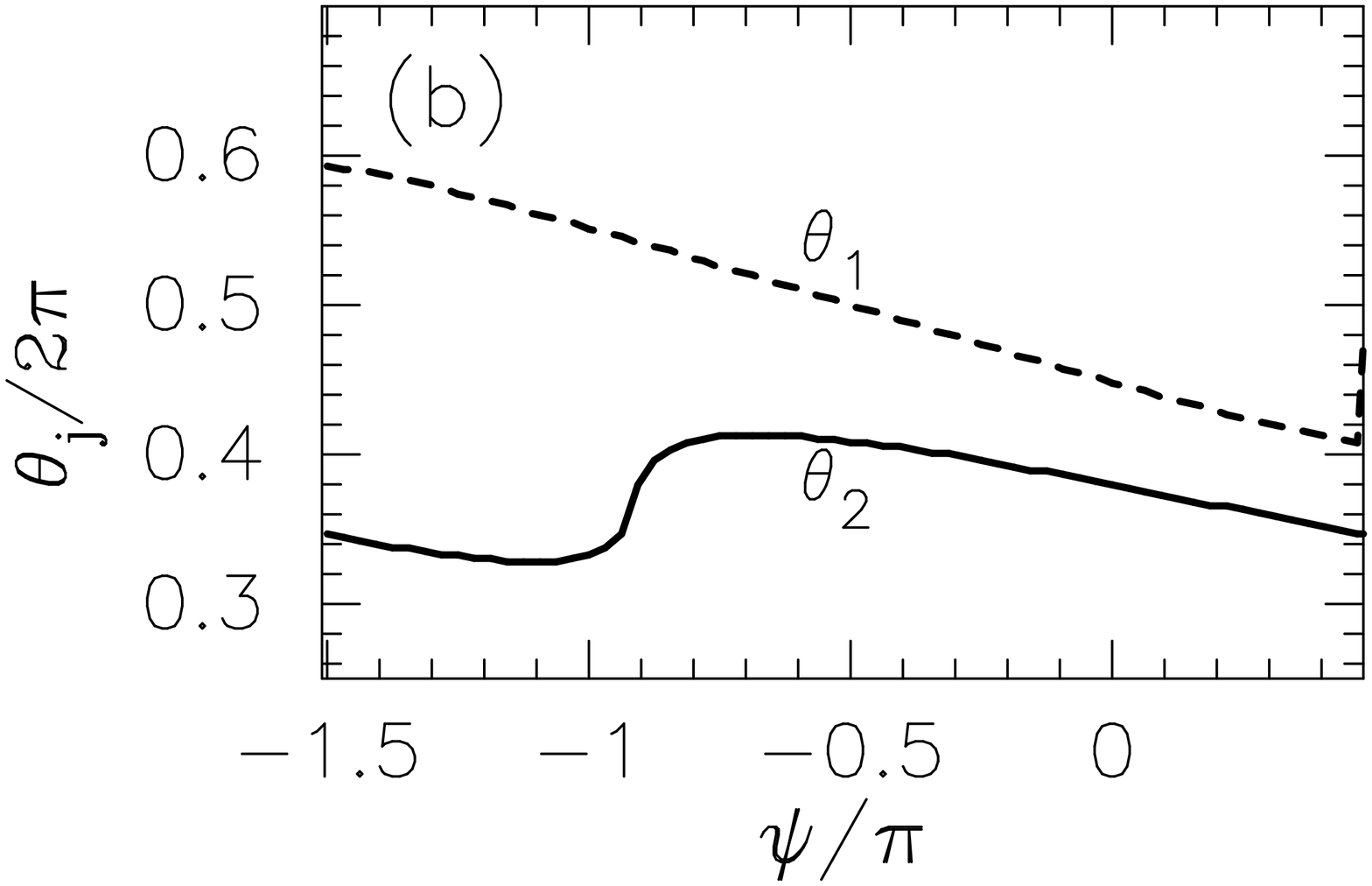}
\caption{
The effect 
 of different time-lags between fundamental and third harmonic
 on the steepness of the pulse.
The field profile is as in eqn. (\ref{eqn-Eplus3H}), 
 with $A(t)=A_0$ and $\gamma=1/16$.
(a)
  Scaled maximum values of $M_j=\textrm{Max}(dE^j/dt)$
  as a function of third harmonic phase offset $\psi$.
 We can see that the maxima of $dE/dt$ and $dE^2/dt$ occur 
  at different $\psi$.
(b)
  Position on the pulse $\theta_j = \omega_1 t_j$ of 
  the maxima plotted in the top frame.
Note that the kinks in (a) and (b) occur for offsets
 that give the {\em longest} shocking distances, 
 not the shortest.
}
\label{fig-pulsecomparison-shifts}
\end{figure}

The primary purpose of this paper is to understand the principles of
 carrier shock formation in the presence of dispersion.  
Some simple ideas about the role of dispersion 
 can be understood from eqn.(\ref{eqn-Eplus3H}) using
 the insight from eqns.(\ref{eqn-moc1}) -- (\ref{eqn-moc4}), 
 i.e. 
 that the rate of change of $E^2$ is the
 critical factor in shock development, 
 rather than that of $E$.
If only the fundamental and third
 harmonic components are considered (as in eqn.(\ref{eqn-Eplus3H})), 
 the effective refractive indices are 
~
\begin{eqnarray}
  \Delta n_{NL:1} &=& n_1 + n_2 \left( 1 + 3 \gamma^2 \right) I_1
,
\label{eqn-delta-refindex1}
\\
  \Delta n_{NL:3} &=& n_3 + n_2 \left( 3 +  \gamma^2 \right) I_1 
.
\label{eqn-delta-refindex3}
\end{eqnarray}
Evidently, 
 the relative phase velocity of the two waves is affected by
 both linear and nonlinear dispersion, 
 so the phase $\psi$ in eqn.(\ref{eqn-Eplus3H}) will
 vary accordingly as the pulse propagates. 
Fig. \ref{fig-pulsecomparison-shifts}, 
 which shows how $dE^2/dt$ varies with $\psi$, 
 suggests that shocking is likely to be
 exacerbated when $\psi$ is small and positive, 
 but moderated when $\psi$ is negative.  
Broadly speaking, 
 the former case will be promoted by anomalous dispersion 
 and the latter by normal dispersion; 
 however, 
 the process is clearly complicated, 
 since it involves time-dependent phase
 shifts between the waves 
 and the interplay of linear and nonlinear dispersion.   
Of course, 
 since carrier shocking relies on the establishment and
 maintenance of specific phase relationships between a set of harmonics,
 it must be expected that strong dispersion of either sign will disrupt
 the shock formation process.  
On the other hand, 
 the simple argument that has been offered suggests that 
 shocking may be tolerant to a degree of anomalous dispersion, 
 but not to a similar amount of normal dispersion.  
In general, 
 therefore, 
 a graph of shocking signature versus refractive index mismatch 
 might be expected to exhibit a shock region where 
 nonlinearity dominates dispersion 
 (displaced in the direction of anomalous dispersion), 
 surrounded by a shock-free region where dispersion dominates 
 the nonlinearity.  
Moreover, 
 if a dominant coherence length $L_C$ can be defined, 
 it is reasonable to expect shocking to occur when this exceeds
 the characteristic SPM length ($L_{SPM}$).
As we shall see, 
 all these features are borne out by the numerical results.

\begin{figure}[ht]
\includegraphics[width=0.80\columnwidth,angle=-0]{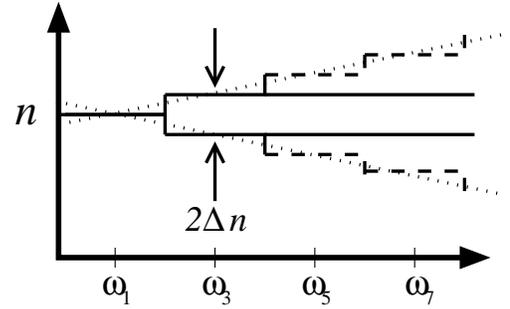}
\caption{
Types of refractive index profile used.
The {\em solid} lines show a single refractive index step midway between 
 $\omega_1$ and $\omega_3$, 
 so that the fundamental is always phase mismatched from its
 higher harmonics, 
 although they remain perfectly phase matched 
 with each other.
The {\em dashed} lines show multiple refractive index steps, 
 each of the same size, 
 and always midway between subsequent harmonics.
The {\em dotted} lines show a linear refractive index gradient
 which gives the same mismatch between subsequent harmonics
 as the multi-stepped case.
We do not show lines for the case of a single step at $4\omega_1$
 to avoid cluttering the figure.
}
\label{fig-dispersiondiag}
\end{figure}

In our numerical simulations, 
 we make extensive use of model dispersion curves, 
 which enable refractive index differences between harmonics
 to be freely controlled, 
 and lead to uncomplicated boundaries between the 
 shock and no-shock regimes.  
The dispersion profiles shown in fig. \ref{fig-dispersiondiag} 
 duly contain either refractive index steps $\Delta n$ 
 at the midpoints between successive odd harmonics, 
 or smooth gradients $\delta$ that provide a similar net change.
In the simplest option, 
 there is a single step at $2\omega_1$, 
 in which case the dominant coherence length is clearly 
 $L_C = \pi / \left| k_3 - 3 k_1 \right|
 = \pi c / 3 \omega_1  \left| n_3 - n_1 \right|$;
 we have also tried a single step at $4\omega_1$, 
 which has a shorter $L_C$.
In all cases, 
 when $n$ increases with frequency, 
 the situation corresponds broadly to normal dispersion, 
 while decreasing $n$ corresponds to anomalous dispersion.   
The step size (or gradient) is chosen
 to be comparable to values in 
 fused silica at $\omega_1 = 1.5$rad/fs, 
 where the refractive index differences between the lower harmonics are 
 $\Delta n_{1,3} \approx 0.06$ and 
 $\Delta n_{3,5} \approx 0.12$\cite{Gilles-MV-1999pre}.
While we consider the case of fused silica itself in 
 Section \ref{S-applications},
 the results based on the model dispersion characteristics are
 invaluable for understanding the essential principles 
 of carrier shock formation in dispersive media.

%
\section{Results and discussion}
\label{S-results}

We will now analyze our 
 numerical simulations of carrier shocking in 
 the presence of dispersion
 on the basis of the principles discussed above.
Results are included for both a CW wave 
 and for a single-cycle pulse.
The CW case gives slightly wider shocking regions,
 but the differences are minor.
This is because the effect of the pulse envelope 
 on the field amplitudes and gradients of the central 
 carrier oscillation is small, 
 except when considering sub-cycle pulses.
In the results we present, 
 the energy conservation and LDD measures 
 reveal the presence of sharp boundaries between 
 the shocking and non-shocking regions.

\begin{figure}[ht]
\begin{center}
\includegraphics[height=0.62\columnwidth,angle=-90]{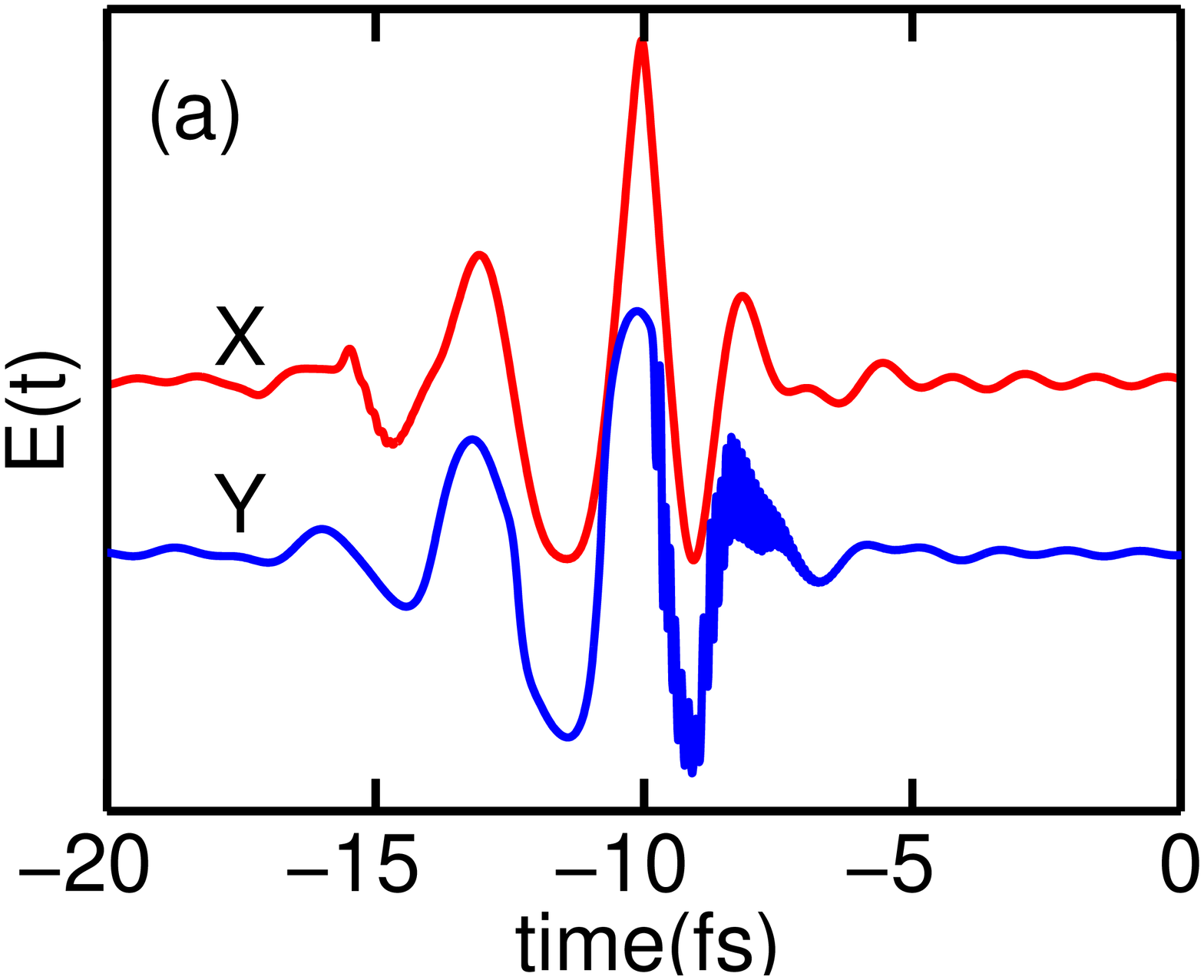}~~~~~~~~~\\
~\includegraphics[height=0.80\columnwidth,angle=-90]{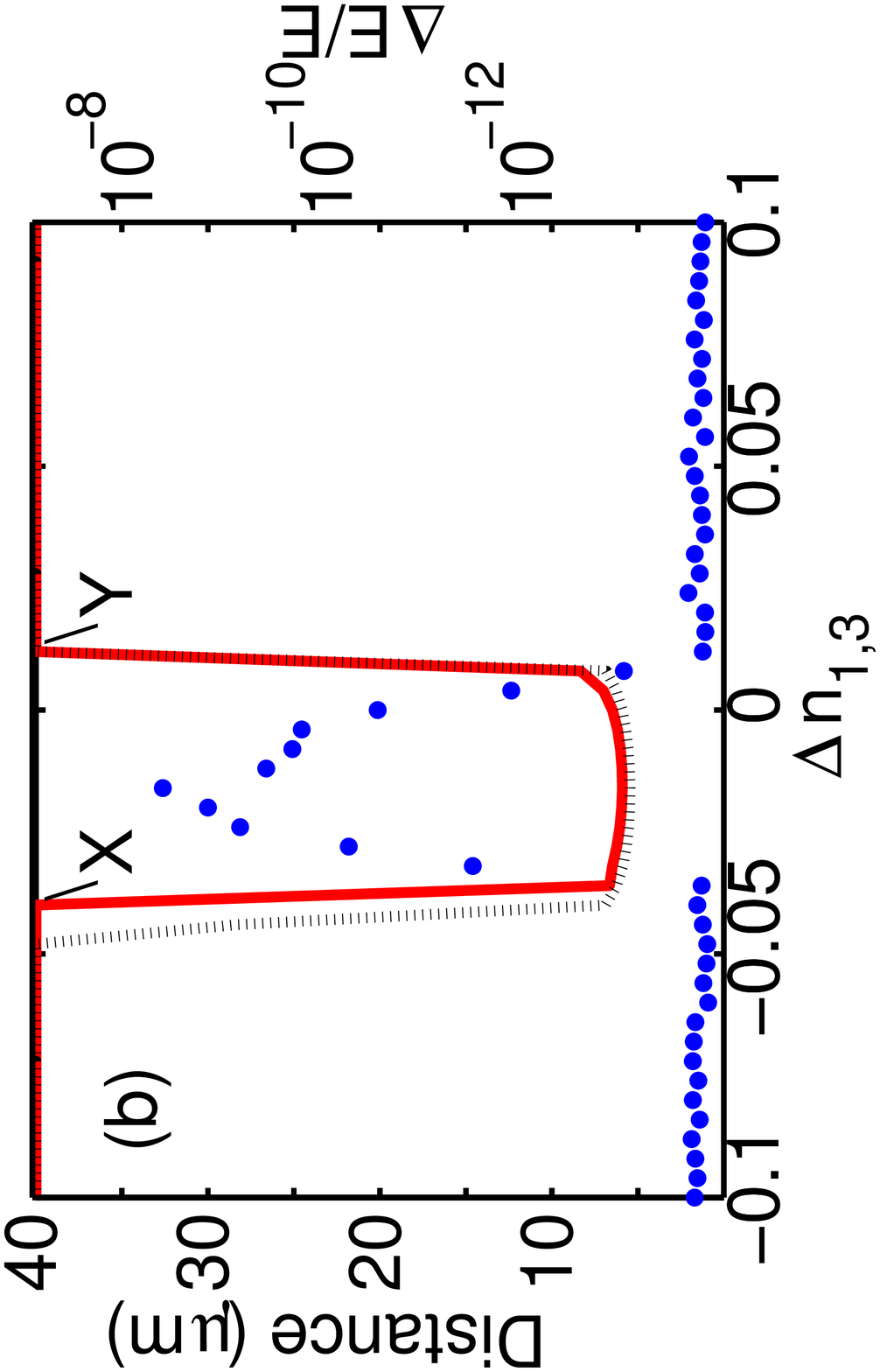}
\end{center}
\caption{
 Carrier shocking for CW and pulsed ($\tau=0.9\dot{3}$) cases with
 a single refractive index step at $2\omega_1$.
Frame (a) compares chirped pulses at a 40$\mu$m propagation distance 
 from the non-shock region
 immediately outside of the shocked region;
 the upper curve (X) is for the negative step, 
 the lower (Y) for the positive step.  
Frame (b) shows the correlation 
 in the region where $L_{C} \lesssim L_{SPM}$
 between 
 energy conservation failure 
 ({\em logarithmic} right hand scale, dots) 
 and the LDD detected shocking distance
 (left hand scale, solid line).
The dotted line shows the LDD results for a CW field ($\tau=\infty$).
}
\label{fig-CWS-singlestep13}
\end{figure}

\begin{figure}[ht]
\begin{center}
\includegraphics[height=0.80\columnwidth,angle=-90]{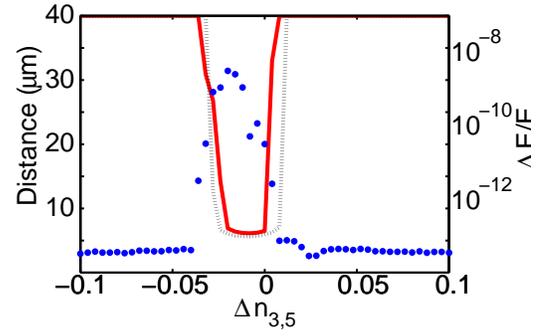}
\end{center}
\caption{
 Carrier shocking for CW and pulsed ($\tau=0.9\dot{3}$) cases with
 a single refractive index step at $4\omega_1$;
showing the correlation
 in the region where $L_{C} \lesssim L_{SPM}$
 between 
 energy conservation failure 
 ({\em logarithmic} right hand scale, dots) 
 and the LDD detected shocking distance
 (left hand scale, solid line).
The dotted line shows the LDD results for a CW field ($\tau=\infty$).
}
\label{fig-CWS-singlestep35}
\end{figure}

\begin{figure}[ht]
\begin{center}
\includegraphics[height=0.80\columnwidth,angle=-90]{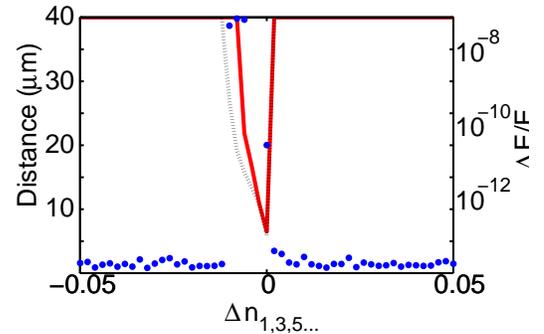}
\end{center}
\caption{
 Carrier shocking for CW and pulsed ($\tau=0.9\dot{3}$) cases with
 multiple refractive index steps;
showing the correlation
 in the region where $L_{C} \lesssim L_{SPM}$
 between 
 energy conservation failure 
 ({\em logarithmic} right hand scale, dots) 
 and the LDD detected shocking distance
 (left hand scale, solid line).
Note the narrower range of $\Delta n$ as compared to the 
 previous two graphs.
The dotted line shows the LDD results for a CW field ($\tau=\infty$).
}
\label{fig-CWS-multistep}
\end{figure}

Results for single refractive index steps
 at $2\omega_1$ and $4\omega_1$ are presented in 
 figs. \ref{fig-CWS-singlestep13} and \ref{fig-CWS-singlestep35}
 respectively,
 whereas
 fig. \ref{fig-CWS-multistep} has a step midway between all harmonics.
In both single stepped cases, 
 a useful coherence length can be defined on 
 the basis of the mismatch between the fundamental and the harmonic
 just above the step.
In the multi-stepped case, 
 there is no easy way to define a dominant coherence length.

Fig. \ref{fig-CWS-singlestep13}(a) shows 
 the pulse profiles in the anomalous (negative) 
 and normal (positive) single step cases, 
 for the smallest step size at which shocking did {\em not} occur.
The first obvious difference between the profiles is 
 the opposite walk-off direction of the third harmonic, 
 although the third harmonic contribution is hard to see in the anomalous case.
The second is that the pulse profiles exhibit
 distinctly different characteristics according to the sign of 
 the dispersion.  
Narrow spikes are visible in the anomalous case, 
 whereas profiles with more rounded maxima occur for normal dispersion.
A possible interpretation is suggested by 
 fig. \ref{fig-pulsecomparison-shifts}, 
 where we saw that anomalous dispersion tends to 
 create regions of higher gradient.

The shocking region in
 fig. \ref{fig-CWS-singlestep35} has a similar outline to that of
 fig. \ref{fig-CWS-singlestep13}(b)
 except that it is slightly narrower, 
 as expected from the coherence length discussion above.  
However, 
 the shocking region in 
 fig. \ref{fig-CWS-multistep} is much narrower, 
 especially given the change in the $\Delta n$ scale.
This is because the multi-stepped nature of the refractive index
 leads to a correspondingly large range of coherence lengths, 
 with shorter ones corresponding to those spanning several steps. 
Since the shocking region in the multi-step case has 
 reduced in size by a factor of two or three, 
 a reasonable inference might be that the 
 dominant coherence length 
 results from interaction over 
 two or three refractive index steps.

\begin{figure}[ht]
\begin{center}
 \includegraphics[height=0.80\columnwidth,angle=-90]{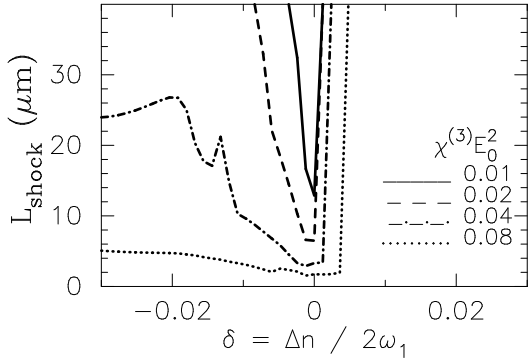}
\end{center}
\caption{
Shocking distance vs mismatch for weak refractive index 
 gradients $\delta$, 
 and a range of nonlinearities.
This shows an abrupt cut-off at positive $\Delta n$,
 but a relatively gradual one at negative $\Delta n$.
The initial pulses are identical to those in 
 fig. \ref{fig-CWS-singlestep13}.
No CW results are shown, 
 since the dispersion experienced by the field 
 would be identical to the multi-stepped case shown in fig. \ref{fig-CWS-multistep}.
}
\label{fig-CWS-linear}
\end{figure}

In realistic media, 
 the refractive index will vary smoothly with frequency, 
 and the group velocity can different from the phase velocity.
We can approximate this situation 
 most simply using a refractive index gradient
 rather than a series of steps.
The results using the LDD method in this case can be seen on 
 fig. \ref{fig-CWS-linear},
 where now we also vary 
 the strength of the nonlinearity.
As in the previous cases, 
 we see a well defined shocking regime that is asymmetric about 
 the non-dispersive case.

Detailed examination shows that 
 the curves in fig. \ref{fig-CWS-linear}
 exhibit a marked similarity, 
 and can be brought into near perfect coincidence by applying 
 the scaling $\chi^{(3)} \rightarrow \chi^{(3)}/m$, 
 $L \rightarrow m L$, and $\Delta n \rightarrow \Delta n / m$.
We also get a comparable similarity for each of 
 figs.
 \ref{fig-CWS-singlestep13},\ref{fig-CWS-singlestep35},\ref{fig-CWS-multistep},
 when simulations at nonlinear strengths of
 $\chi^{(3)}E_0^2 = 0.01$ and $0.04$ are added to the results.
This demonstrates that the character of the shocking is 
 dominated by the sign of inter-harmonic phase velocity differences, 
 not by the local group velocity dispersion at each harmonic.
Thus anomalous (normal) dispersion is primarily interesting 
 because it gives a negative (positive) refractive index shift
 between successive harmonics.
We leave the complicated (and far more subtle) effects 
 of group velocity differences 
 or dispersion
 for later work.

An important feature of all the results is 
 the pronounced asymmetry,  
 with shocking persisting much further 
 into the anomalous dispersion regime; 
 we have even seen it for 
 the case of a weakly parabolic refractive index.

We can quantify this asymmetry by considering 
 linear and nonlinear 
 contributions to the phase matching of the harmonics
 (i.e. linear and nonlinear refractive index shifts),
 as described at the beginning of section \ref{S-dispersive}.
In the simple single step case shown on
 fig. \ref{fig-CWS-singlestep13}(b), 
 the 1st--3rd harmonic phase shift will dominate,
 because it applies to the two most intense spectral components.
We can calculate the SPM-induced refractive index shift between 
 the fundamental and third harmonic 
 with eqns. (\ref{eqn-delta-refindex1}) and (\ref{eqn-delta-refindex3})
 to be 
 $\Delta n_{NL:1} - \Delta n_{NL:3} = -0.011$, 
 since $\chi^{(3)} E^2 = 0.02$ at the 
 peak of the carrier oscillations, 
 the refractive index is $n_0^2=2$, 
 and $2 n_2 E_1^2 = 2 \times 3 \chi^{(3)} E^2 / 8 n_0$.
This is roughly comparable to the 
 offset of the shocking region, 
 which is centred at about $\Delta n_{1,3} \simeq -0.014$.
We cannot expect perfect agreement, 
 since the calculations ignore 
 the role of higher harmonic generation and 
 depletion of the fundamental.
Whilst a simple calculation is reasonably successful 
 in this single-step case, 
 it cannot be applied for a realistic medium -- 
 or indeed to the situation shown in 
 fig. \ref{fig-CWS-linear}.
There,
 the effects of dispersion,
 higher harmonic generation, 
 and nonlinear refractive index shifts are inextricably intertwined.

To summarize,
  we have demonstrated that shocking is strongly
 dependent on the interplay between $L_{C}$ and $L_{SPM}$, 
 with shorter $L_{C}$'s (increasing $\Delta n$'s) 
 decreasing the likelihood of shocking.
We have also deduced the reasons for 
 the strong {\em asymmetry} of the shocking region.

%
\section{Applications and experiments}
\label{S-applications}

In sections \ref{S-introduction} and \ref{S-shockdetection}, 
 we discussed how imminent carrier shocking 
 might be recognised computationally.
In considering whether shocking might be detectable experimentally, 
 we must now decide how it might manifest itself in 
 the laboratory. 
A mathematical discontinuity is clearly not a physical possibility, 
 even if Rosen \cite{Rosen-1965pr} did manage to accommodate it theoretically,
 albiet at the expense of energy conservation.
In practice, 
 the increasing field gradients (and spectral broadening) 
 that precede a shock will inevitably engender new 
 physical processes that will limit the steepening.
Indeed,
 we have already seen this happening 
 in the previous two sections, 
 where dispersion has been seen to frustrate the self-steepening process; 
 the next barrier would be 
 the time-scale of the nonlinear response.

\begin{figure}[ht]
\begin{center}
\includegraphics[height=0.90\columnwidth,angle=-90]{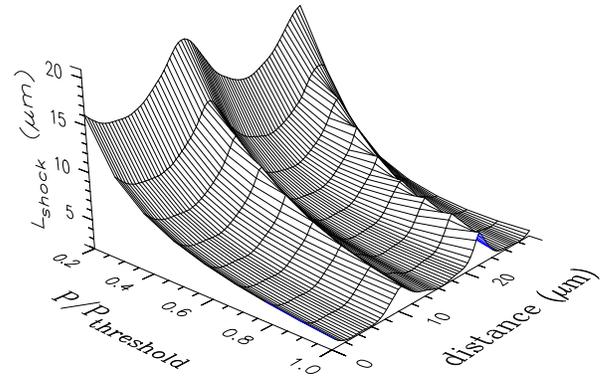}
\caption{
Numerically predicted shocking distances $L_{shock}$ in fused silica.
These are the MOC predicted distances for the waveform, 
 assuming the dispersion was (abruptly) neglected, 
 the shortest distance shown is about 1.9$\mu$m.
The damage threshold for fused silica was taken to be 
$P_{threshold}=50$TW/cm$^2$.
} 
\label{fig-fusedsilica}
\end{center}
\end{figure}

%

Although there is no question of a mathematical discontinuity 
 being observed in an experiment, 
 the recent advances in the measurment of optical pulse profiles 
 (see e.g. \cite{Goulielmakis-UKBYSWKHDK-2004sci}),
 suggest that it might be possible to observe carrier steepening.
Indeed, 
 it has already been predicted \cite{Gilles-MV-1999pre} that 
 noticeable steepening effects 
 could occur for pulses propagating in fused silica.
Unfortunately, 
 in our own simulations of this process, 
 the LDD shock detection was not triggered at any realistic pulse intensity.
Although visible steeping did occur, 
 at no point was a seriously distorted waveform approached, 
 and the effects would have been milder if we had 
 included the finite response time of the nonlinearity.
The results presented in fig. \ref{fig-fusedsilica} show 
 the steepest gradient recorded as a function of distance 
 and pulse intensity for these simulations.
The third harmonic coherence length for these parameters
 is about 7$\mu$m, 
 and we can attribute 
 the regular variation with distance seen in the figure to 
 the third harmonic component aligning with 
 successive oscillations of the fundamental
 as the waves move across each other.
The key reason why incipient shocking was not detected is because the pulse
 frequency lies in a region of weak normal dispersion, 
 whereas we have shown in section \ref{S-dispersive} that 
 a region of weak anomalous dispersion (assuming it could be achieved)
 would be favourable.

The current interest in media
 with tailored dispersion characteristics
 \cite{PhotonicBandgapMaterials,PhotonicCrystals,MicrostructuredFibres}
 raises the interesting question of whether it might be possible 
 to engineer a material that maximize self-steepening.
A major stumbling block would be the need for 
 control over many harmonic orders; 
 however, 
 for a narrow-band pulse, 
 only the dispersion characteristics close to the harmonics would be relevant, 
 which might perhaps make the technical challenge less formidable.

The recognition that dispersion control can enhance (or reduce) 
 carrier self-steepening suggests other applications if we
 widen our horizons to encompass the 
 more general idea of carrier {\em shaping}. 
In this case, 
 we would exploit both nonlinearity and dispersion control 
 to optimize the shape of the carrier oscillations 
 for a particular experiment. 
Applications such as high harmonic generation 
 (e.g. \cite{Christov-MK-1997prl}) might well benefit from
 suitably designed carrier wave modulation.

%
\section{Conclusions}
\label{S-conclusions}

In this paper 
 we have investigated carrier shock formation, 
 developed criteria for detecting its onset in numerical simulations, 
 and shown how it is influenced by a range of parameters, 
 particularly dispersion.
We have also obtained remarkable agreement between  
 numerical simulations and theoretical predictions of the 
 shocking distance in the dispersionless limit, 
 and shown that the process is sensitive to both 
 CEP and pulse duration.

Although we have confirmed that 
 shocking occurs in a narrow parameter range, 
 this is far from being the whole picture.
In particular, 
 there is a distinct asymmetry between the 
 anomalous and normal dispersion regimes.
The former leads to shocking signatures such as the 
 appearance of narrow spikes
 as the higher harmonics interfere on 
 the steepening part of the pulse profile.
In contrast, 
 normal dispersion creates no such features, 
 and the pulse profiles have a rather blunt appearance. 
The asymmetry arises from the 
 effect of the nonlinear refractive index on 
 the dispersion induced phase mismatch.

The conclusion to be drawn from our results is clear:
 if they could be engineered, 
 materials with wide regions of weakly anomalous dispersion are
 much better candidates for generating steep, 
 shock-like field profiles
 than those (such as silica) with 
 a weak normal dispersion.
Detecting incipient shock formation in materials
 like fused silica is likely to be a near impossible task, 
 given the constraints imposed by their damage thresholds.

%

\end{document}